\documentclass[tighten, times, twocolumn]{aastex631}
\shorttitle{Magnetic Reconnection as the Driver of the Solar Wind}
\shortauthors{Raouafi et al. 2023}
\graphicspath{{./}{figures/}}

\begin{document}

\title{Magnetic Reconnection as the Driver of the Solar Wind}

\author[0000-0003-2409-3742]{Nour E. Raouafi}
\affiliation{The Johns Hopkins Applied Physics Laboratory, Laurel, MD 20723, USA}

\author[0000-0001-8480-947X]{G. Stenborg}
\affiliation{The Johns Hopkins Applied Physics Laboratory, Laurel, MD 20723, USA}

\author[0000-0002-0494-2025]{D. B. Seaton}
\affiliation{Southwest Research Institute, Boulder, CO 80302, USA}

\author[0000-0002-5233-565X]{H. Wang}
\affiliation{Institute for Space Weather Sciences, New Jersey Institute of Technology, University Heights, Newark, NJ 07102, USA }
\affiliation{Big Bear Solar Observatory, New Jersey Institute of Technology, Big Bear City, CA 92314, USA}
\affiliation{Center for Solar-Terrestrial Research, New Jersey Institute of Technology, University Heights, Newark, NJ 07102-1982, USA}

\author[0000-0001-5099-8209]{J. Wang}
\affiliation{Institute for Space Weather Sciences, New Jersey Institute of Technology, University Heights, Newark, NJ 07102, USA }
\affiliation{Big Bear Solar Observatory, New Jersey Institute of Technology, Big Bear City, CA 92314, USA}
\affiliation{Center for Solar-Terrestrial Research, New Jersey Institute of Technology, University Heights, Newark, NJ 07102-1982, USA}

\author[0000-0002-7164-2786]{C. E. DeForest}
\affiliation{Southwest Research Institute, Boulder, CO 80302, USA}

\author[0000-0002-1989-3596]{S. D. Bale}
\affiliation{Physics Department, University of California, Berkeley, CA 94720, USA}
\affiliation{Space Sciences Laboratory, University of California, Berkeley, CA 94720, USA}

\author[0000-0002-9150-1841]{J. F. Drake}
\affiliation{Department of Physics, University of Maryland, College Park, MD 20742, USA}

\author[0000-0002-5871-6605]{V. M. Uritsky}
\affiliation{Catholic University of America, 620 Michigan Avenue NE, Washington, DC 20061, USA}
\affiliation{Heliophysics Science Division, NASA Goddard Space Flight Center, Greenbelt, MD 20771, USA}

\author[0000-0002-6975-5642]{J. T. Karpen}
\affiliation{Heliophysics Science Division, NASA Goddard Space Flight Center, Greenbelt, MD 20771, USA}

\author[0000-0002-4668-591X]{C. R. DeVore}
\affiliation{Heliophysics Science Division, NASA Goddard Space Flight Center, Greenbelt, MD 20771, USA}

\author[0000-0003-1281-897X]{A. C. Sterling} 
\affiliation{NASA/Marshall Space Flight Center, Huntsville, AL 35812, USA}

\author[0000-0002-7572-4690]{T. S. Horbury} 
\affiliation{The Blackett Laboratory, Imperial College London, London, SW7 2AZ, UK}

\author[0000-0001-9457-6200]{L. K. Harra}
\affiliation{PMOD/WRC, Dorfstrasse33 7260 Davos Dorf, Switzerland}
\affiliation{ETH-Zurich, H{\"{o}}nggerberg campus, HIT building, Z{\"{u}}rich, Switzerland}

\author[0000-0002-2358-6628]{S. Bourouaine}
\affiliation{The Johns Hopkins Applied Physics Laboratory, Laurel, MD 20723, USA}

\author[0000-0002-7077-930X]{J. C. Kasper}
\affiliation{BWX Technologies, Inc., Washington DC 20002, USA}

\author[0000-0001-6289-7341]{P. Kumar}
\affiliation{Department of Physics, American University, Washington, DC 20016, USA }
\affiliation{Heliophysics Science Division, NASA Goddard Space Flight Center, Greenbelt, MD 20771, USA}

\author[0000-0002-6924-9408]{T. D. Phan} 
\affiliation{Space Sciences Laboratory, University of California, Berkeley, CA 94720, USA}

\author[0000-0002-2381-3106]{M. Velli} 
\affiliation{Earth Planetary and Space Sciences, UCLA, CA 90095, USA}



\begin{abstract}

We present EUV solar observations showing evidence for omnipresent jetting activity driven by small-scale magnetic reconnection at the base of the solar corona. We argue that the physical mechanism that heats and drives the solar wind at its source is ubiquitous magnetic reconnection in the form of small-scale jetting activity (i.e., a.k.a. jetlets). This jetting activity, like the solar wind and the heating of the coronal plasma, are ubiquitous regardless of the solar cycle phase. Each event arises from small-scale reconnection of opposite polarity magnetic fields producing a short-lived jet of hot plasma and Alfv\'en waves into the corona. The discrete nature of these jetlet events leads to intermittent outflows from the corona, which homogenize as they propagate away from the Sun and form the solar wind. This discovery establishes the importance of small-scale magnetic reconnection in solar and stellar atmospheres in understanding ubiquitous phenomena such as coronal heating and solar wind acceleration. Based on previous analyses linking the switchbacks to the magnetic network, we also argue that these new observations might provide the link between the magnetic activity at the base of the corona and the switchback solar wind phenomenon. These new observations need to be put in the bigger picture of the role of magnetic reconnection and the diverse form of jetting in the solar atmosphere.

\end{abstract}



\section{Introduction} \label{sec:intro}

Solar and stellar winds are ubiquitous flows of charged particles (i.e., electrons, protons, and heavier ions) permeating the astral spheres \citep{1962Sci...138.1095N}. Through these winds, stars lose angular momentum, slow down their rotation as they age, shape planetary systems, and affect the composition and the physical and chemical evolution of planetary atmospheres and, consequently, the habitability of these planets \citep{luftinger_gudel_johnstone_2014,2017AA...597A..14G}. How the solar wind is generated at the source, heated, and accelerated, and what determines its variability, are long-standing fundamental questions.

The genesis of the hot and highly dynamic plasma in the corona and the solar wind is among astrophysics’ most challenging and long-standing questions. The solar wind has three main regimes: fast, slow, and transient. The fast solar wind, with speeds typically over 500 km/s, originates from the interior of coronal holes (i.e., open magnetic-field regions). The source of the slow wind is highly debated \citep{2016SSRv..201...55A}. It apparently arises from the interfaces between closed-field regions, such as active regions and quiet Sun, and the edges of open-field coronal holes \citep{2015ApJ...805...84D}. The fast solar wind is less dense and hotter than the slow wind, and has photospheric composition, whereas the slow wind has coronal composition. At 1 AU, the fast wind is mainly Alfv\'enic , whereas most slow wind is more variable and non-Alfv\'enic \citep{1991AnGeo...9..416G,2005LRSP....2....4B,2019Natur.576..237B,2019Natur.576..228K,2022ApJ...932L..13B}, although uncommon streams of Alfv\'enic slow wind have been reported \citep{2015ApJ...805...84D,1981JGR....86.9199M,2019MNRAS.483.4665D,2020A&A...633A.166P}. Two major theories have been proposed to explain the solar wind's genesis via heating and acceleration: magnetic reconnection \citep{1988ApJ...330..474P,1992sws..coll....1A,2003JGRA..108.1157F} and magnetohydrodynamic (MHD) wave turbulence \citep{1971JGR....76.3534B}. Transients, such as coronal mass ejections (CMEs), are considered a third solar-wind regime that drives space weather and correlates with the sunspot cycle \citep{raouafi_solar_2021}.

Jetting in the solar atmosphere manifests in different forms (e.g., spicules [\citealt{1968SoPh....3..367B,1972ARA&A..10...73B,2000SoPh..196...79S,2007PASJ...59S.655D}], jets [\citealt{1992PASJ...44L.173S,2016SSRv..201....1R}], and surges [\citealt{1996ApJ...464.1016C}]). There is growing evidence that this jetting plays a key role in supplying the corona and the solar wind with mass and momentum, and may provide enough energy to power the solar wind \citep{2007Sci...318.1574D,2011Natur.475..477M,2014Sci...346A.315T}. Coronal jet signatures have been traced out to several Mm in X-ray/extreme ultraviolet (EUV) observations, up to several solar radii in white-light images \citep{1998ApJ...508..899W}, and beyond 1 AU in {\emph{in~situ}} measurements \citep{2006ApJ...639..495W,2008ApJ...675L.125N,2012ApJ...750...50N}. At lower altitudes, \citet{2007PASJ...59S.655D}  identified a similar phenomenon in the chromosphere, the Type~II spicules, which are typically observed in the chromospheric Ca~II 854.2 nm and H$\alpha$ lines \citep{2009ApJ...705..272R}, and are heated as they propagate upward. However, there is no evidence for Type II spicules reaching coronal temperatures and altitudes as coronal plumes and jets do. Observations from the IRIS \citep{2014SoPh..289.2733D} and {\emph{SDO}} missions suggest that the spicular cool plasma falls back to the solar surface \citep{2015ApJ...815L..16S}. Whether spicules contribute to the solar wind and how much is not well known. A recent study by \citet{2022ApJ...937...71S} suggests that significantly more spicules than observed were needed to drive the solar wind.

Close to the Sun, Parker Solar Probe ({\emph{PSP}}) measurements reveal a highly structured solar wind dominated by high-amplitude Alfv\'en waves. The magnetic field is often observed to rotate over $90^\circ$ forming reversals or switchbacks \citep{2019Natur.576..237B,2019Natur.576..228K}, which were observed before by {\emph{Ulysses}} \citep{1999GeoRL..26..631B}, {\emph{WIND}} \citep{2011ApJ...737L..35G}, and {\emph{Helios}} \citep{2018MNRAS.478.1980H}. They were, however, scarce at large heliodistances. These switchbacks also occur in patches separated by quiet periods where the field is nearly radial. Several reports discuss the potential origins of these structures, which can be put in two categories: coronal origin \citep{2020ApJ...894L...4F,2020ApJ...896L..18S,2021A&A...650A...2D} and {\emph{in~situ}} solar-wind origin \citep{2020ApJ...891L...2S,2020ApJ...902...94R,2021ApJ...915...52S,2021ApJ...918...62M,2021ApJ...913L..14H,2021ApJ...909...95S}. \citet{2021ApJ...923..174B} and \citet{2021ApJ...919...96F} found that the scale size of switchback patches correlates with the scale size of supergranules on the solar surface, favoring a coronal origin for the switchbacks. However, how the switchbacks form in the corona and the driving physical mechanism near the solar surface remain unclear. This topic is hotly debated \citep{2020ApJ...896L..18S,2021A&A...650A...2D,2020ApJ...891L...2S,2021ApJ...915...52S,2021ApJ...923..174B}, partly because of the lack of clear observational evidence of the processes responsible for heating and driving the solar wind near the base of the solar atmosphere.

Small-scale jetting, or jetlets, was discovered in coronal plumes in equatorial coronal holes by \citet{2014ApJ...787..118R}. Coronal plumes are bright structures extending from the magnetic network into high coronal altitudes \citep{1997ApJ...484L..75W}. They are particularly prominent in images of  total solar eclipses, and were historically known as coronal rays \citep{1950BAN....11..150V}. Plumes are also observed to extend to solar wind altitudes \citep[i.e., $\sim45~R_\odot$;][]{2001ApJ...546..569D}. They are brighter but cooler than surrounding interplumes regions observed as darker (i.e., lower density) lanes in EUV and white-light images of the solar corona. For further details on coronal plumes and jets, see the reviews by \citet{2011A&ARv..19...35W} and \citet{2016SSRv..201....1R}. \citet{2014ApJ...787..118R} showed that the small-scale and high-frequency jetting (i.e., jetlets) at the base of coronal plumes is driven by interchange magnetic reconnection, and that it sustains them for long periods of time \citep[see also][]{2018ApJ...868L..27P,2019ApJ...887L...8P,2021ApJ...907....1U,2022ApJ...933...21K}.

Here we show evidence that ubiquitous jetting at tiny scales (a few hundred km) driven by interchange magnetic reconnection near the base of the corona could be the origin of the heating and acceleration of the solar wind. We interpret the magnetic field switchbacks \citep{2019Natur.576..237B,2019Natur.576..228K} as tracers of this small-scale explosive magnetic activity.

\section{Ubiquitous jetting activity at the base of the solar corona} \label{JettingCorona}

\begin{figure*}[ht!]
\begin{center}
\includegraphics[width=0.75\textwidth]{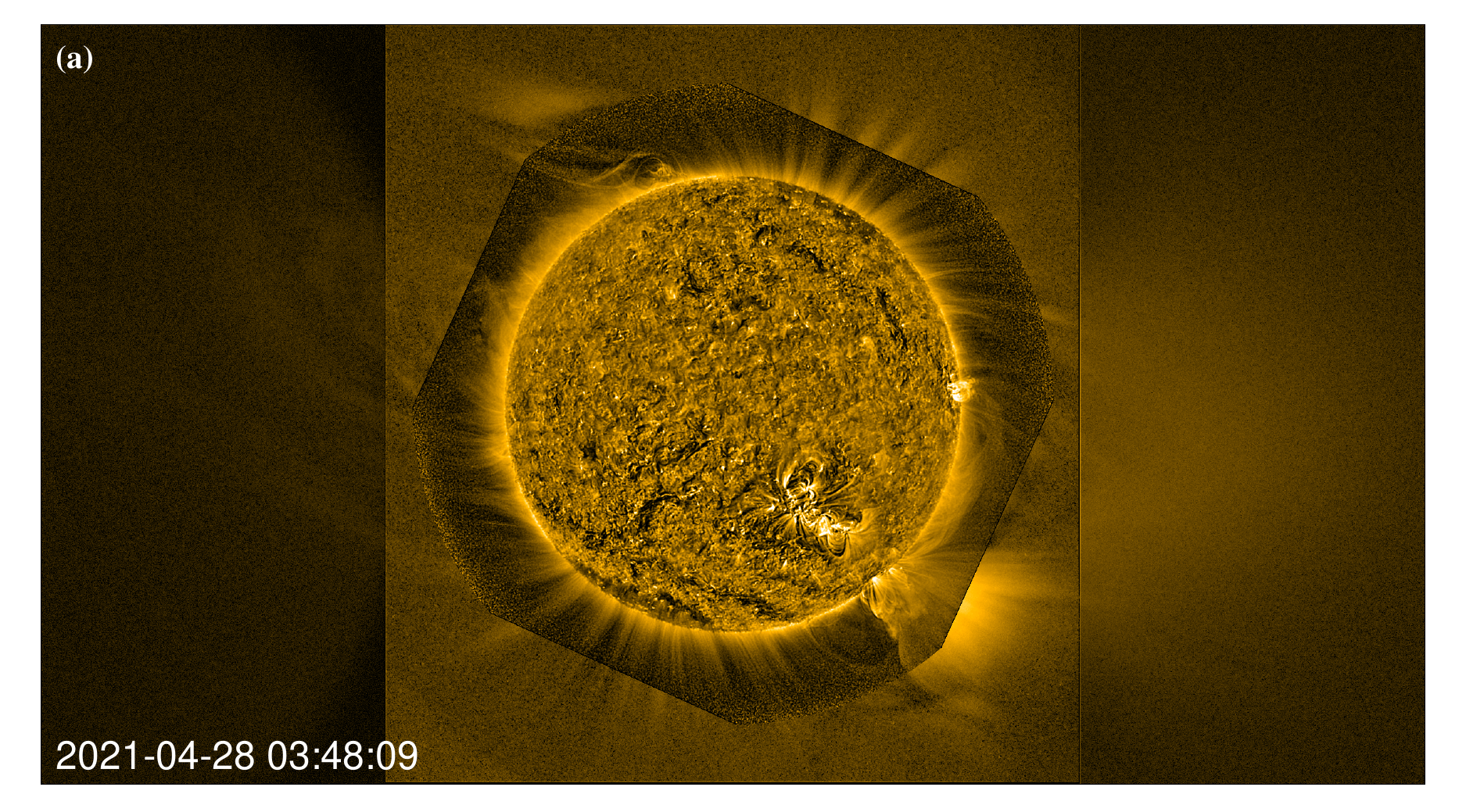} \\
\includegraphics[width=0.75\textwidth]{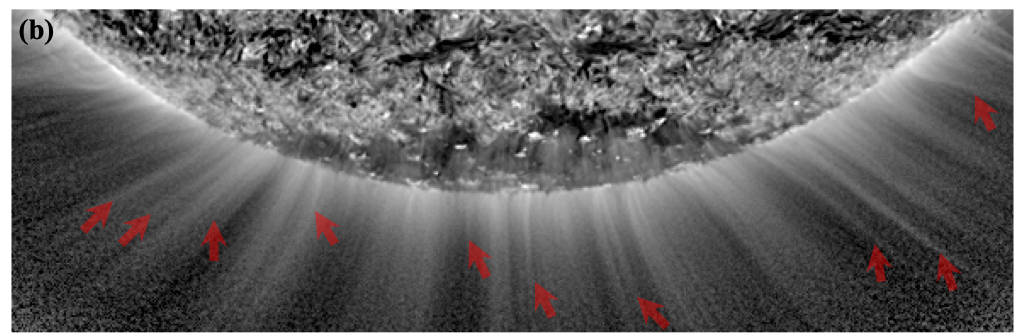}
\caption{(a) Composite of {\emph{SDO}}/AIA and {\emph{GOES-R}}/SUVI 171~{\AA} images showing the small-scale activity at the base of the solar corona and its extension to higher altitudes (see movies in the Supplemental material). The maximum extent of the jetlets in the AIA field of view is limited by the instrument sensitivity. Estimates of their occurrence rate and size are also limited by the temporal and spatial resolution of the instrument. The SUVI image maps the structures observed at the coronal base into the solar wind. The accompanying movies illustrate the highly dynamic and continuous nature of this phenomenon. (b) AIA image (171~{\AA}) showing the jetlet structures as elongated features above the solar polar limb. Examples of jetlet events are indicated by the arrows.  \label{FigAIASUVI}}
\end{center}
\end{figure*}

The high-resolution, high-cadence observations from space missions such as {\emph{SOHO}} \citep{1995SSRv...72...81D}, {\emph{Hinode}} \citep{2007SoPh..243....3K}, {\emph{STEREO}} \citep{2008SSRv..136....5K}, {\emph{SDO}} \citep{2012SoPh..275....3P}, and {\emph{SolO}} \citep{2020A&A...642A...1M} show tremendous diversity of multi-scale explosive activity ranging from enormous flares \citep{2011LRSP....8....6S} and CMEs \citep{2011LRSP....8....1C} down to bright-point eruptions \citep{2019LRSP...16....2M} and coronal jets \citep{1992PASJ...44L.173S,2016SSRv..201....1R}. Solar observations suggest that magnetic reconnection plays a predominant role in the evolution of these structures by enabling the impulsive conversion of stored magnetic energy to plasma kinetic and thermal energy and to nonthermal particles. In contrast to sunspots and active regions that nearly disappear at the minimum of the solar cycle, small-scale activity (e.g., jets, bright points, etc.) is omnipresent regardless of the solar-cycle phase \citep[see, e.g.,][]{2014ApJ...792...12M,2019LRSP...16....2M}. In fact, jetting resulting from magnetic reconnection is evidently a fundamental process on the Sun. Reconnection-driven jets are not restricted to the open magnetic-field regions (i.e., coronal holes) but also occur in closed structures, heating the plasma to high temperatures. A particular example of this activity is the tiny jets (i.e., jetlets) observed at the base of plumes within equatorial coronal holes \citep{2014ApJ...787..118R}. Previous analyses were, however, confined to particular coronal structures, namely plumes in equatorial coronal holes \citep{2021ApJ...907....1U,2022ApJ...933...21K} or singular jetlets \citep{2018ApJ...868L..27P,2019ApJ...887L...8P}. Jetlets are minuscule reconnection events between open and closed magnetic flux resulting in collimated plasma ejections into the solar corona \citep{2014ApJ...787..118R,2022ApJ...933...21K,2018ApJ...868L..27P,2019ApJ...887L...8P}. \citep{2014ApJ...787..118R} found that jetlets are the primary driver of coronal plumes sustaining them for days and weeks. \citep{2022ApJ...933...21K} also found quasiperiodic energy releases (equivalent to nanoflare energies, i.e., $10^{24}$~ergs) and associated jetlets at the base of plumes that could contribute significant mass flux to the solar wind.

The mechanism producing jetlets within plumes also occurs elsewhere on the solar disk. A careful analysis of the {\emph{SDO}}/AIA and {\emph{GOES-R}}/SUVI observations reveals that this phenomenon is much more pervasive than merely coronal plumes. Figure~\ref{FigAIASUVI}a and Figure~\ref{FigAIASUVI}b are a composite of AIA and SUVI wide-field images \citep{2021NatAs...5.1029S} and an AIA zoomed view of the southern polar region, respectively. The raw images show hazy structures extending to high coronal altitudes. The processed images using the multi-resolution image-processing technique reveal tiny bursts of hot plasma permeating nearly all coronal structures. The lifetime of these events ranges from tens of seconds to several minutes. The ubiquity and the highly dynamic nature of this activity in the off-limb corona are striking (see the movies provided in the Supplemental material for details). Based on the analysis of long series of continuous SDO/AIA images, we find that this off-limb small-scale activity persists over time, indicating that it extends over the whole solar surface (i.e., coronal holes, the quiet Sun, and active regions; see the movies in the supplemental material).

\section{Magnetic reconnection driving the small-scale jetting} \label{MagRecJet}

\begin{figure}[ht!]
\includegraphics[width=0.475\textwidth]{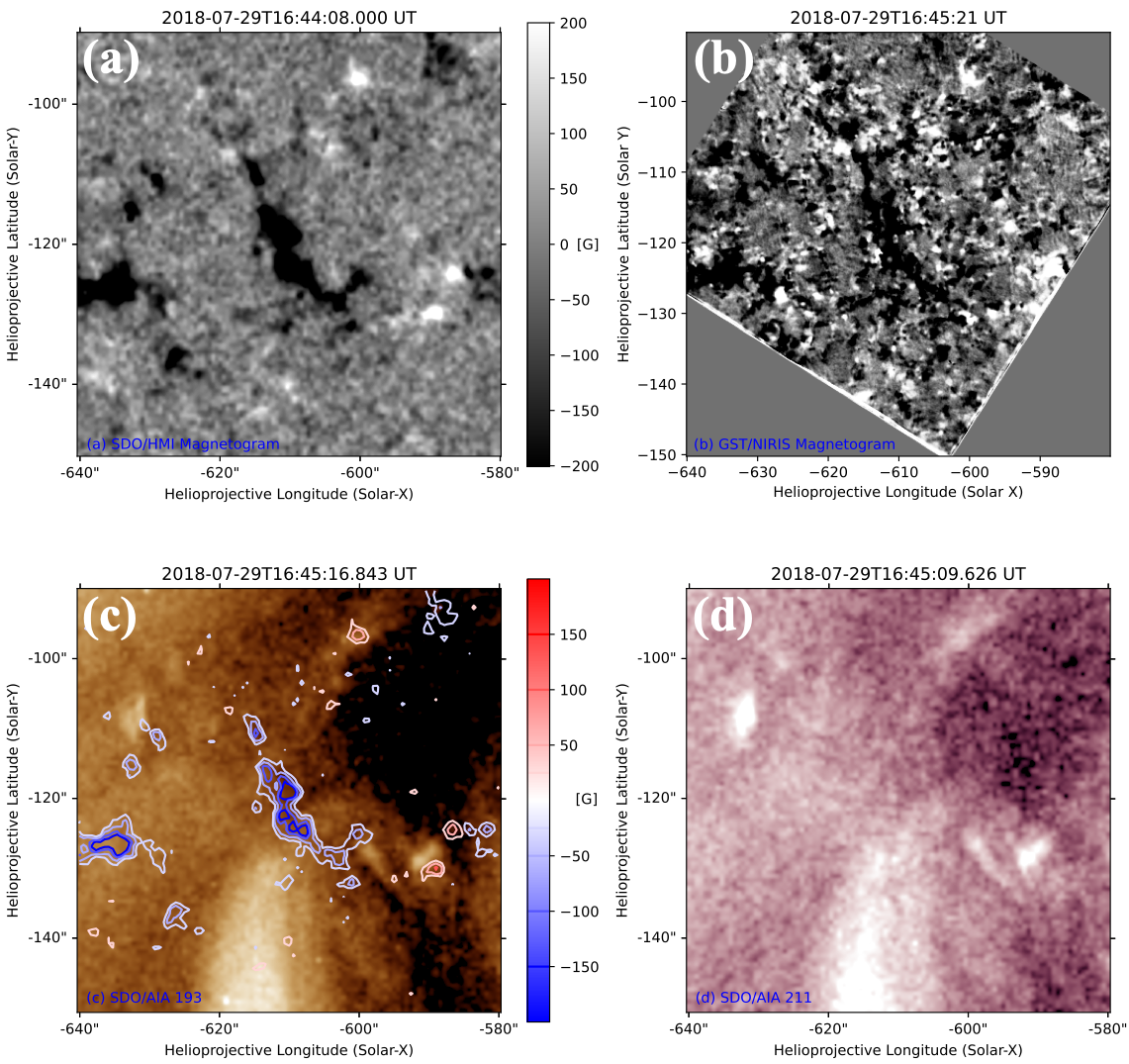}
\caption{Co-temporal magnetograms from the {\emph{SDO}}/HMI (a) and {\emph{BBSO}}/GST-NIRIS  (b) instruments with respective spatial resolutions of $1\arcsec$ and $0\arcsec.2$. The magnitude NIRIS magnetograms is scaled to the HMI unit scale. The displayed magnetic fields saturate at $\pm200$~G. The corresponding EUV images are in the 193~{\AA} (c) and 211~{\AA} (d) channels of SDO/AIA, respectively. Red (blue) contours represent positive (negative) network fields and magnetic elements. Panel (a) shows that only strong-field regions are resolved at low resolution, and most of the solar disk area seems to be void of any significant flux. The magnetogram images change dramatically with increasing spatial resolution and instrument sensitivity (panel (b)). In particular, apparent void regions and unipolar patches show significantly more mixed polarities, a favored landscape for magnetic reconnection.  \label{FigBBSOGST}}
\end{figure}

Figure~\ref{FigBBSOGST}a,b displays magnetograms from the {\emph{SDO}} Helioseismic and Magnetic Imager \citep[HMI;][]{2012SoPh..275..207S} and the 1.6~m Goode Solar Telescope (GST) at the Big Bear Solar Observatory (BBSO). The HMI and GST magnetograms have a spatial resolution of $1\arcsec$ and $0\arcsec.2$, respectively. Only relatively strong field regions can be observed at the low resolution and sensitivity of HMI, with hints of a diffuse opposite polarity. This clear view is primarily due to two instrumental factors: the polarimetric sensitivity and the low resolution that leads to the Zeeman-cancellation of opposite-polarity fields within the resolution element. The magnetic-field landscape changes dramatically by improving the instrumental sensitivity and increasing the spatial resolution. Most unipolar flux concentrations observed with low-resolution instruments become fragmented at high resolution, as can be seen clearly in the GST sub-arcsecond magnetograms. A multitude of multi-scale magnetic elements of highly mixed polarity are present throughout the instrument’s field of view. Magnetic reconnection between background network magnetic fields (at the supergranule boundaries, the base of the open fields in coronal holes) and opposite-polarity intranetwork fields are likely the cause of small-scale jetting. This comports with the finding that switchbacks have a modulation scale of supergranules \citep{2021ApJ...923..174B,2021ApJ...919...96F}. An animated sequence of GST magnetograms is provided in the Supplemental material.

In the GST magnetograms\footnote{For more details on how the BBSO/GST magnetograms were produced and the magnetic fine structures were identified, see \citet{2022ApJ...924..137W} and Appendix A.}, a significant number of magnetic bipoles appeared in the regions devoid of magnetic flux at coarser resolution. These small-scale, highly mixed polarity fields are a rich medium for magnetic reconnection. During about 90 minutes of continuous observations, 1434 cancellation events were identified in the quiet Sun/coronal-hole boundary region in the GST $70\arcsec\times70\arcsec$ field of view. Most notably, the distribution of these sites seems to be uniform as there is no appreciable difference between the quiet Sun and the coronal hole (Figure~\ref{FigBBSOGST}c,d), an indicator of the universality of small-scale reconnection in the lower solar atmosphere. The magnetic flux-cancellation rate in the observation is $1–2\times10^{18}$~ Mx~Mm$^{-2}$~hr$^{-1}$. 88 cancellations were associated with H$\alpha$ spicules, of which 61 were rooted in network field concentrations presumably open to the solar wind. Among them, 7 produced detectable EUV jetlets above the spicules. Assuming that these occurrences are typical of those over the whole Sun, scaling the observed frequencies yields about 600 flux-cancellation events~s$^{-1}$ generating about 35 H$\alpha$ spicules~s$^{-1}$ and 3 EUV jetlets~s$^{-1}$. We expect that such cancellation/reconnection sites would produce additional eruptive events below the smallest currently observable scales.

Magnetic reconnection also generates Alfv\'enic perturbations (waves, fronts, and shocks). The simultaneous generation of the radial flows associated with jetlets and Alfv\'enic perturbations is a natural consequence of reconnection \citep{2017ApJ...834...62K,2017ApJ...837..123U,2018ApJ...866...14R}. \citet{2007Sci...318.1580C} showed evidence for Alfv\'en waves in solar X-ray jet. These Alfv\'enic waves are crucial for heating and accelerating the solar wind plasma, and for generating turbulent flows at higher coronal altitudes \citep{2011ApJ...743..197C}. 

\subsection{Coronal Jetting Rate, Mass and Energy Fluxes} \label{CorJetRates}

\begin{figure}[ht!]
\includegraphics[width=0.475\textwidth]{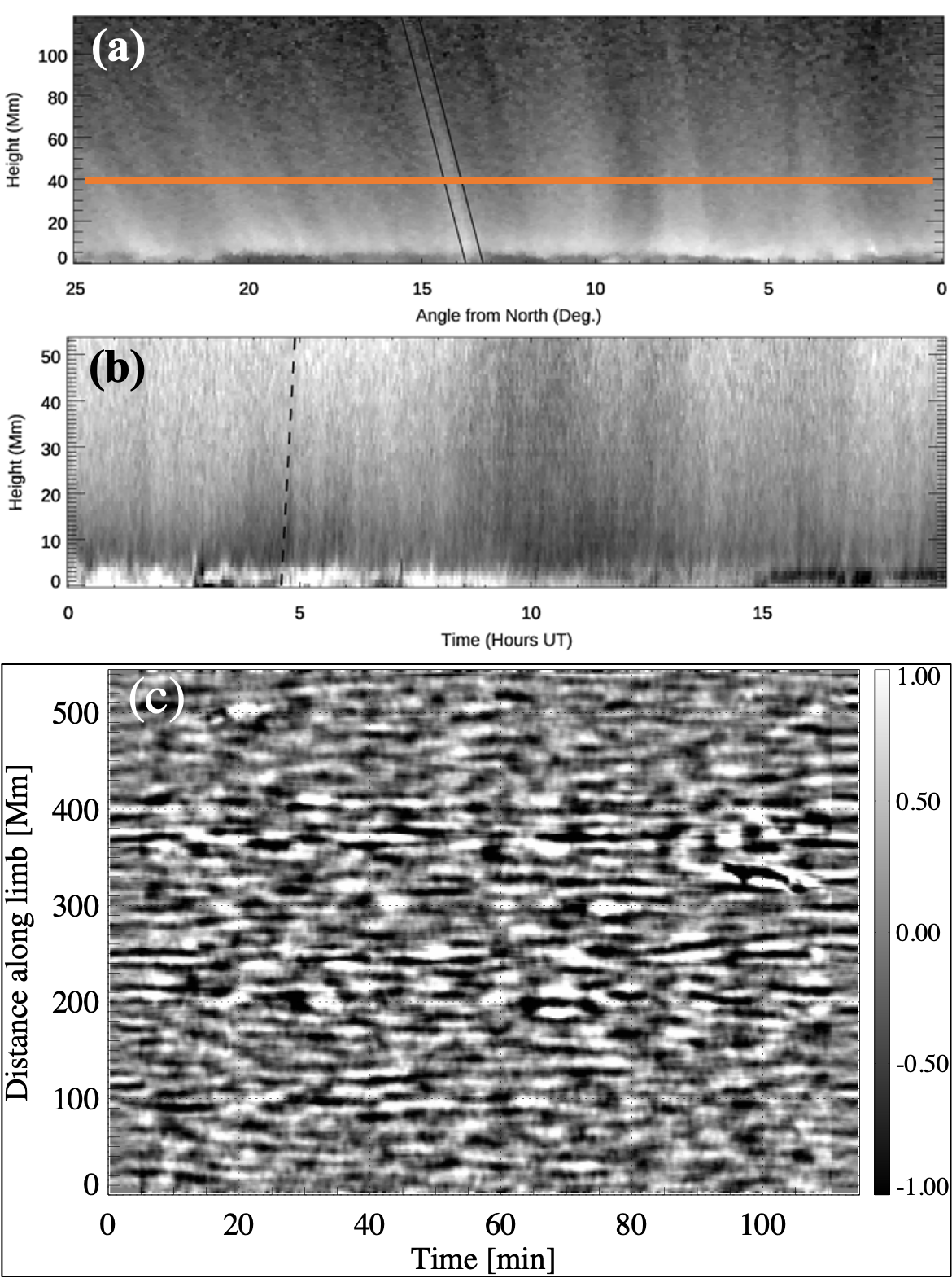}
\caption{(a) Polar projection of AIA 171~{\AA} images of the northern solar polar region showing coronal bright (dark) structures (e.g., plumes, interplume regions, etc.). (b) Time-distance diagram along the virtual slit marked by the dark lines in (a). Jetlets emanating from small-scale magnetic reconnection persist at all times and dominate the activity at the base of the solar corona. (c) Diagram showing the coronal structures crossing the orange virtual slit in (a) as a function of time at an altitude of 40 Mm above the solar polar limb. The jetlets are the bright structures.  \label{FigJetRate}}
\end{figure}

Figure~\ref{FigJetRate}a shows an {\emph{SDO}}/AIA 171 Å polar projection of the northern polar region ($\pm25^\circ$). The two black lines mark the area used to build the time-distance diagram in Figure~\ref{FigJetRate}b. The jetlet events along the virtual slit show as streaks whose slopes yield an average speed of $\sim150$~km~s$^{-1}$. Figure~\ref{FigJetRate}c shows the signal in the EUV image crossing a circular slit at an altitude of 40 Mm above the solar polar limb (i.e., orange line in panel (a)) as a function of time. The jetlets are the bright structures with typical lifetimes of several minutes. To estimate the jetlets crossing the artificial slit (orange line in Figure~\ref{FigJetRate}a), we apply a Fourier analysis that gives us the jetting rate. We then inverted this rate into the total number of jetlets over the several-hour period we considered for this analysis. Our analysis of these two diagrams provides an occurrence rate of $\sim2500$ jetlets per day over a position-angle interval of $50^\circ$ above the northern polar cap. This jetlet rate is orders of magnitude higher than the reported $\sim160$ X-ray jets per day per solar hemisphere \citep{2007PASJ...59S.771S,2015A&A...579A..96P}, even though our analysis underestimates the jetlet occurrence rate by about 30\%. The jetlet detection is also limited by the instrument sensitivity and the spatial and temporal resolution of the data. The X-ray jets are, in contrast, significantly larger and longer-lived, and individually more energetic. 

The jetlets typically have a width of $2\arcsec-3\arcsec$ \citep{2022ApJ...933...21K}, a speed of ~150 km~s$^{-1}$, and a lifetime of $5-10$ minutes. Assuming a coronal density of $5\times10^8$~cm$^{-3}$ at the base of these events, the particle ejection rate into the corona resulting from each small-scale jetlet is about $7.5\times10^15$~cm$^{-2}$ s$^{-1}$. This amounts to about $3\times10^{32}$~protons~s$^{-1}$ and $1\times10^{35}$~protons total over the lifetime of the jetlet, assuming that all of the ejected plasma escapes (hence this value should be regarded as an upper limit). To account for the entire solar wind loss of $6\times10^{35}$~protons~s$^{-1}$ \citep{2000JGR...10527217V,2016ApJ...833L..21W,2020ApJ...904..199W} requires roughly $2\times10^3$ jetlets to be active at any instant and 6 jetlets~s$^{-1}$ to be initiated over the full Sun. This last number is comparable to the rate extrapolated from the BBSO/GST measurements discussed above.

The kinetic energy injected into the corona by each jetlet is $1.2\times10^6$ erg~cm$^{-2}$ s$^{-1}$ or $5\times10^{22}$~erg~s$^{-1}$, assuming the same jetlet width as quoted above.  If $2\times10^3$ jetlets are active at any instant, the total jetlet kinetic-energy injection rate is $1\times10^{26}$~erg~s$^{-1}$. By comparison, the overall solar kinetic-energy loss rate (assuming an asymptotic flow speed of 500 km~s$^{-1}$) is about $1\times10^{27}$ erg~s$^{-1}$. Clearly the injected jetlet plasma must be accelerated further by the coronal thermal pressure plus wave pressure to reach the asymptotic wind speed. 

The BBSO/GST analysis shows that the magnetic flux density of the reconnecting bipoles at the photosphere is $\sim190$~G. At coronal altitudes, we estimate the strength of the reconnecting field to lie in the range $5-10$~G. The magnetic energy released to the plasma during the reconnection process is assumed to be partitioned between plasma bulk flow and heating. (see the Supplemental material for the detailed calculations.)

The total solar jetlet-generation rate of $\sim6$ jetlets~s$^{-1}$ (i.e., $5\times10^5$ jetlets~d$^{-1}$) greatly exceeds the estimated  $\sim0.03$ jetlets~s$^{-1}$ (i.e., $2.5\times10^3$ jetlets~d$^{-1}$) estimated from the limb observations with {\emph{SDO}}/AIA. However, the latter encompassed just $1/{2\pi}$ of the solar circumference, and it was restricted to jetlets within a narrow, undetermined angle $\delta$ from the limb onto and behind the disk. The full-Sun and limb-detected results are consistent for $\delta\approx0.03$~rad, equivalent to a linear distance $d\approx20$~Mm from the limb.

All of these rates fall in the ranges required to drive the solar-wind plasma at the base of the corona in the quiet Sun and coronal holes. Thus, our analysis supports our contention that the ubiquitous, small-scale jetting activity (jetlets) driven by magnetic reconnection can account for essentially all of the mass and energy lost by the Sun to the solar wind.

A critical aspect of measuring the reconnection-driven jetting at the base of the corona is the dependence on the resolution of the magnetic-field data and the EUV images. We expect that magnetic field data with significantly higher resolution and polarization accuracy, for instance from the 4-m Daniel K. Inouye Solar Telescope (DKIST), would provide a substantially higher incidence of reconnection events, resulting in a considerably higher jetting rate. This will, of course, affect the estimated heating and acceleration of the coronal solar-wind plasma.

\section{Parker Solar Probe Observations and their Connection to the Corona}


Using Ulysses’ fast solar wind measurements above the solar poles ($>1$ AU), \citet{1995JGR...10023389N} showed that the so-called micro-streams, where the solar wind speed deviates by $>20$ km~s$^{-1}$ from the average, are of solar origin. Historically, micro-streams were thought to be related to coronal plumes, although this relationship cannot fully explain their properties. \citet{2012ApJ...750...50N} argued that micro-streams are related to episodic rather than quasi-stationary sources. Based on the work by \citet{2008ApJ...682L.137R}, which found a causal relationship between jets and plumes, \citep{2012ApJ...750...50N} confirmed that the micro-streams are of solar origin, and their properties can be explained if the fast ones result from jetting activity at the base of the corona.

\begin{figure}[ht!]
\includegraphics[width=0.475\textwidth]{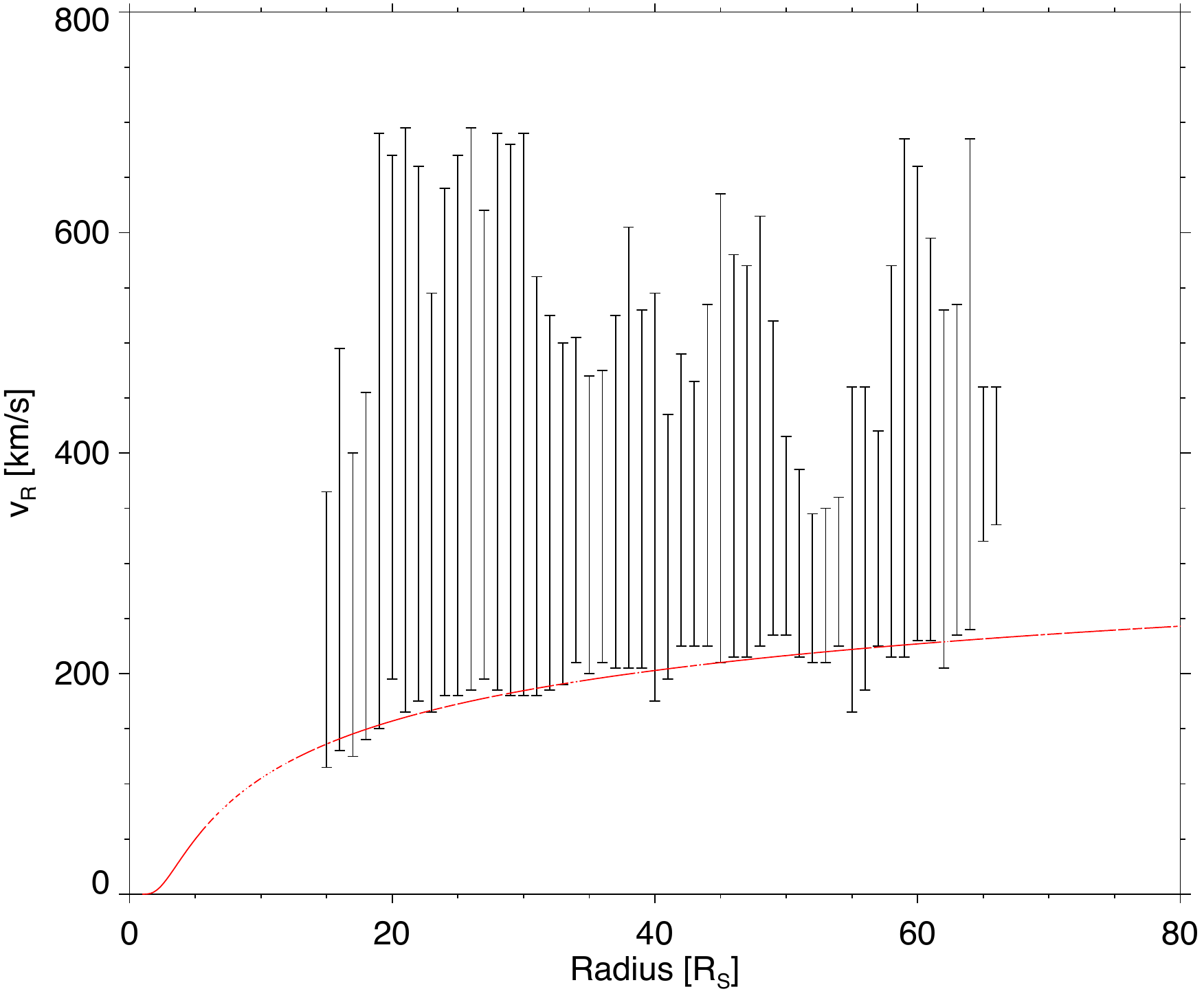}
\caption{The vertical bars show the spread of solar wind velocities measured by PSP as a function of helio-distance during the first ten encounters. The red curve, which bounds the measurements on the lower end, seems to indicate that the base solar wind behaves like the Parker model \citep{1958ApJ...128..664P}. Above that boundary, plasma jets dominate the solar wind, which may suggest tracers in the solar wind of the coronal jetting activity.    \label{FigSWJets}}
\end{figure}

Closer to the Sun, {\emph{PSP}} observed predominantly Alfv\'enic solar wind (both fast and slow) during its perihelion encounters ($<0.25$ AU). Streams of non-Alfv\'enic flows have been reported, but they represent only a very small fraction of the observations. The prevalence of Alfv\'enic flows in the inner solar wind is an important clue about the nature of the solar wind as it emerges from its source(s). During the first perihelion encounter , a small equatorial hole was identified as the source of the observed slow wind \citep{2019Natur.576..237B}. For the other encounters, the models show that the spacecraft is frequently connected magnetically to the edges of the polar coronal holes and their equatorial extensions. At the base of the solar atmosphere, only remote-sensing data are available to assess the physical processes that might occur at the origin of the solar wind flow. Current data quality is much higher than in previous decades, so with {\emph{PSP}} flying so close to the Sun, linking the {\emph{in~situ}} measurements to remote-sensing observations at much lower altitudes in the solar atmosphere is an exciting possibility. 

The data also show that the solar wind speed, as measured by {\emph{PSP}}, is dominated by radial-velocity jets \citep{2019Natur.576..228K} superimposed on the background Parker-like wind (see Figure~\ref{FigSWJets}). The vertical bars represent the spread of the speeds of these plasma jets. The red curve marks the low bound of these speeds, remarkably resembling a Parker-like solar wind. The structuring of the inner solar wind and the dominance of {\emph{in~situ}} plasma jetting may also indicate the signature of the small-scale magnetic reconnection and jetting at the base of the solar corona. Magnetic reconnection produces Alfv\'enic waves that eventually make their way to high altitudes and whose dissipation heats and accelerates the solar wind plasma \citep[see, e.g.,][]{2007Sci...318.1574D,2011Natur.475..477M}.

Switchbacks are short magnetic field rotations that are ubiquitously observed in the solar wind. They are consistent with local folds in the magnetic field rather than changes in the magnetic connectivity to solar source regions \citep{2019Natur.576..237B,2019Natur.576..228K}.
The large number of the omnipresent eruptive jetting events observed at the base of the corona is a credible explanation of the source of the magnetic-field switchbacks observed by {\emph{PSP}}. The jetlets are the direct product of ubiquitous magnetic reconnection at small spatial and temporal scales. The EUV images from {\emph{SDO}} and {\emph{GOES-R}}/SUVI and the magnetic-field data from the BBSO/GST provide clear evidence for the preponderance of small-scale reconnection at these sites. The EUV images, particularly in the solar polar regions, show a semi-regular spacing of brighter and darker coronal structures. The brighter areas exhibit a much higher jetlet occurrence than the darker ones. Hence, the switchback patchiness could be explained by the varying magnetic connection of the spacecraft to sites with different susceptibilities to reconnection events. The quiet periods (dark in EUV) would correspond to locations with lower event rates, while the times of strong connectivity to regions with higher event rates characterized by more jetlets/switchbacks. 

Different models have been suggested to explain the formation of switchbacks: (1) interchange reconnection \citep[e.g.,][]{2020ApJ...894L...4F,2020ApJ...896L..18S,2021ApJ...913L..14H,2021A&A...650A...2D,2022ApJ...925..213A}; (2) steepening of Alfv\'en waves and/or Alfv\'enic turbulence \citep{2020ApJ...891L...2S,2021ApJ...918...62M,2021ApJ...915...52S}; (3) due to roll up from nonlinear Kelvin-Helmholtz instabilities 
\citep{2020ApJ...902...94R}; and (4) through magnetic field lines that stretch between sources of slower and faster wind \citep{2021ApJ...909...95S}.
We postulate that the omnipresent magnetic reconnection and the resulting jetting in the corona satisfy most if not the proposed switchback models. Magnetic reconnection produces impulsive plasma jets and Alfv\'en waves, which are the principal inputs for most switchback models (e.g., Alfv\'en waves, shear flows, etc.).

\section{Conclusions. How is the solar wind born?}

The coronal holes where the fast solar wind originates are regions of open magnetic fields. \citep{1999Sci...283..810H} used coronal observations in the Ne$^{7+}$ 770~{\AA} spectral line to find evidence for strong outflows coinciding with the boundaries of the chromospheric network. Although they did not discuss the physical mechanism generating these outflows, they suggested that the wind is rooted in the boundaries of this network. \citep{2005Sci...308..519T} suggested that these areas are open to the corona and could be the source of the wind. The present data show the predominance of intermittently driven hot plasma outflows at small scales. These jetlets are omnipresent, much like the solar wind, regardless of the phase of the sunspot cycle. Evidence for reconnection in the low solar atmosphere is present across the entire solar disk, particularly at the boundaries of the chromospheric network (i.e., supergranules). Although these areas are typically dominated by unipolar fields, high-resolution magnetic-field measurements show the presence of minority-polarity intrusions (i.e., the salt and pepper fields) actively moving amongst and canceling with the dominant polarity field to drive the jetlets. 

We believe that magnetic activity at small scales plays the dominant role in shaping the solar atmosphere, heating the corona, and driving the solar wind. We believe the jetlets analyzed here are part of a whole spectrum that extends to much smaller scales. With higher spatial and temporal resolution and greater instrumental sensitivity, therefore, we expect to detect more frequent signatures of magnetic reconnection at finer scales. For instance, DKIST will provide observations with spatial resolution three times better than BBSO/GST. With these data, we expect to identify significantly more fine-scale magnetic reconnection sites providing hot and impulsive plasma jets to the corona and the solar wind, with significant implications for coronal heating and solar-wind acceleration.

Our proposed scenario applies most obviously to the fast solar wind. This originates in coronal hole regions, where the magnetic field is open. Therefore, jets on coronal hole open fields have a direct route to the heliosphere, and therefore can explain the fast solar wind in a straightforward manner.  In contrast, the origin of the slow solar wind is not yet clear, but there is evidence that it originates in closed-field regions and/or at the boundaries between open- and closed-field regions.  Because jets/jetlets occur in the close-field regions also, we expect that they are source of not only the fast wind, but also the slow wind too. Further analyses are, however, required to confirm this. We hope to clarify these points with future PSP observations. 

One crucial aspect of PSP measurement close to the Sun is that almost all the observed solar wind is highly Alfv\'enic. The observed Alfv\'enicity of the wind seems independent of the wind regime. It might indicate a common physical process at the origin of the solar wind and that the difference between the slow and fast wind might result from evolution at higher altitudes. \citet{1990ApJ...355..726W} suggested that super-radial expansion of the coronal magnetic field can generate a slow solar wind such as at the boundaries of coronal holes \citep[see also][]{2019ApJ...873...25P}. Future PSP measurements, during the upcoming closest perihelia together with Solar Orbiter and DKIST observations, hold promise for confirming the links between small-scale magnetic activity and the solar wind, hopefully by inferring direct connections between small-scale reconnection or other magnetic events and small-scale structures in the solar wind.

\begin{acknowledgements}
We are grateful to Dr. Valentin Martinez Pillet for the constructive comments and suggestions, which helped improve the quality of the paper. 

Parker Solar Probe was designed, built, and is now operated by the Johns Hopkins Applied Physics Laboratory as part of NASA’s Living with a Star (LWS) program (contract NNN06AA01C). Support from the LWS management and technical team has played a critical role in the success of the Parker Solar Probe mission. 

SDO is the first mission to be launched for NASA’s Living With a Star (LWS) Program. The SDO/AIA and SDO/HMI data are provided by the Joint Science Operations Center (JSOC) Science Data Processing (SDP). 

Solar UltraViolet Imager (SUVI) product development, analysis, calibration, validation, and data stewardship by CIRES-affiliated authors within National Centers for Environmental Information (NCEI) was supported by National Oceanic and Atmospheric Administration cooperative agreement no. NA17OAR4320101.  

We gratefully acknowledge the use of data from the Goode Solar Telescope (GST) of the Big Bear Solar Observatory (BBSO). BBSO operation is supported by US NSF AGS-1821294 grant and New Jersey Institute of Technology. GST operation is partly supported by the Korea Astronomy and Space Science Institute and the Seoul National University.
\end{acknowledgements}


\appendix

\section{BBSO/GST magnetic field data.}

Taking advantage of high-order correction by the adaptive optics system with 308 sub-apertures \citep{2010AN....331..636C} and the solar speckle interferometric data-reconstruction technique \citep{2008A&A...488..375W}, the observation during $\sim$16:34 – 18:38 UT achieved diffraction-limited resolution under a favorable seeing condition. There is a $\sim20$-minute observation gap between 18:07 – 18:27 UT due to bad seeing. Spectroscopic polarization measurements of Fe I 1.56~$\mu$m were taken by NIRIS with a $0\arcsec$.24 resolution and a 42 s cadence. 

Due to weak polarization signal in the quiet-Sun regions, line-of-sight (LOS) magnetograms are reduced by summing Stokes-V profiles from GST observations to enhance the SNR. The magnetic field strength is scaled the contemporal HMI magnetic-field measurements. The small-scale magnetic elements are tracked with Southwest Automatic Magnetic Identification Suite \citep[SWAMIS;][]{2007ApJ...666..576D} based on similarity heuristics across a time series of magnetograms, by which the magnetic cancellation events are detected and their corresponding magnetic fluxes are calculated.

\section{Energy and particle fluxes from magnetic reconnection}
This section estimates the magnetic energy flux resulting from the small-scale reconnection episodes. This energy flux is transferred to the plasma in the form of bulk flows and heating. We start from the observed particle and kinetic energy ejection rates into the corona and wind, using the following average jetlet properties:
\begin{itemize}
\item Transverse scale (i.e., width): $L_J \approx 3\arcsec \approx 2000$ km
\item Speed: $V_J \approx 150$~ km~s$^{-1}$
\item Lifetime: $\tau \approx 5$~min $=300$~s
\item Density: $n \approx 5\times10^8$~cm$^{-3}$
\end{itemize}

The jetlet speed is determined from the time-distance diagram. It is the projected speed on the plane of the sky, which should be considered as a lower limit on the real jetlet speed. For the electron density, we used a typical plume density at the base of the corona. The jetlets may be denser, perhaps by as much as a factor of 4 \citep{2020ApJ...896L..18S}, but we employ the ambient coronal density to be conservative. There are variations by at least a factor of 2 in the width ($L_J$) and lifetime ($\tau$) of the jetlets that decrease or increase our estimated jetlet contributions to the wind and, therefore, the number of jetlets required to drive the entire solar-wind flux. It is not possible to be precise about these contributions beyond a factor of about 2 in either direction. We have endeavored here to demonstrate that conservatively estimated jetlet contributions to the wind are comparable to the total estimated mass and energy fluxes from the Sun. The resulting values are given in the main text of the paper.

The reconnecting magnetic field strength in the corona is much smaller than the measured average photospheric flux density, $B_{ph}\approx100$~G in quiet Sun and slightly higher in the coronal hole boundary region. Because we do not have direct coronal magnetic-field measurements, we must infer the strength of the reconnecting field, $B_R$. We do this by requiring the magnetic energy released by the reconnection to be sufficient to power the jetlet outflow, plus an assumed equivalent amount of plasma heating. For simplicity, we ignore the unknown but plausible contribution of released magnetic energy to accelerated nonthermal particles, which is a very important and well-known consequence of reconnection in large CMEs and flares.

The magnetic reconnection inflow speed, $V_R$, is assumed to be a fraction 0.1 of the Alfv\'en speed, $V_A = B_R/\sqrt{4\pi\rho}$, associated with the reconnecting field strength, $B_R$. This dimensionless reconnection rate (i.e., the inflow Alfv\'en Mach number) is well established from numerical MHD simulations of fast reconnection, including those specifically of reconnection-driven coronal jets \citep{2017ApJ...834...62K}. The reconnection occurs over a transverse scale $L_R$ that defines the width of the reconnection region in the corona, and which may be smaller than the width $L_J$ of the jetlet. For further detail on the theory of magnetic reconnection, see \citet{1993SSRv...65...59L}.

The kinetic energy flux density and total release rate into the bulk outflow are
\begin{eqnarray}
w_{KE} &=& \frac{1}{2} \rho V_J^2 V_J , \\
W_{KE} &=& w_{KE} L_J^2
\end{eqnarray}
whence $w_{KE} \approx 1.2\times10^6$~erg~cm$^{-2}$~s$^{-1}$ and $W_{KE}\approx 5\times10^{22}$~erg~s$^{-1}$. Assuming that half of the magnetic energy is transferred to the plasma in the form of heating, the total magnetic energy release rate during the reconnection, $W_{ME}$, satisfies $W_{ME} = 2\ W_{KE} \approx 1\times 10^{23}$~erg~s$^{-1}$. We have 
\begin{eqnarray}
w_{ME} &=& \frac{1}{4\pi} \ B_R^2 \ V_R = \frac{0.1}{4\pi}\  B_R^2 \ \frac{B_R}{\sqrt{4\pi\rho}} , \\
W_{ME} &=& w_{ME} \  L_R^2, \\
w_{ME} &=& W_{ME}/L^2_R = 2 \  W_{KE}/L^2_R = 2 \  w_{KE} \ L^2_J/L^2_R,
\end{eqnarray}
which can be solved for the reconnection field strength,
\begin{eqnarray}
B_R^3 &=& 80\  \pi^{3/2} \ \sqrt{\rho} \ w_{ME} \\
           &=& 160\  \pi^{3/2} \ \frac{L_J^2}{L_R^2} \ w_{KE}.
\end{eqnarray}

Substituting values from above, we obtain $B_R \approx 5 \ (L_J/L_R)^{2/3}$~G. The implied field strength depends, as is to be expected, on the ratio of characteristic scales in the reconnection region and in the resultant jetlet. The minimum strength is about 5~G, 5\% of the average photospheric flux density of $\sim100$~G, obtained for $L_R = L_J$. This is a reasonable value for the coronal magnetic field. The implied field strength doubles to about 10~G if we assume that $L_R = L_J/3$.

\section{Supplementary material}






The supplementary material package contains five video files and can be downloaded \href{https://sppgway.jhuapl.edu/sites/default/files/webform/dropbox/Raouafi_ApJ_ArXiv_Movies.tar.gz}{HERE}

{\bf{Movie M1:}} Composite of {\emph{SDO}}/AIA and {\emph{GOES-R}}/SUVI 171~{\AA} image sequences showing the small-scale activity at the base of the solar corona and its extension to higher altitudes. The movie begins on April 28th, 2021 at 00:00:09 and ends the same day 09:57:09. Its real-time duration is 8 seconds.

{\bf{Movie M2:}} {\emph{SDO}}/AIA 171~{\AA} image sequence showing the small-scale activity at the base of the solar corona. The movie starts on April 28th, 2021 at 00:00:09 and ends the same day 09:57:09. Its real-time duration is 8 seconds.

{\bf{Movie M3:}} Zoom on the {\emph{SDO}}/AIA 171~{\AA} image sequence of the northern solar polar region where small-scale activity is clearly visible. The movie starts on April 28th, 2021 at 00:00:09 and ends the same day 09:57:09. Its real-time duration is 8 seconds.

{\bf{Movie M4:}} Zoom on the {\emph{SDO}}/AIA 171~{\AA} image sequence of the southern solar polar region where small-scale activity is clearly visible. The movie starts on April 28th, 2021 at 00:00:09 and ends the same day 09:57:09. Its real-time duration is 8 seconds.

{\bf{Movie M5:}} Animation of high-resolution magnetograms from the BBSO/GST-NIRIS instrument showing the highly-dynamic magnetic fields at small-scales. The number of the cancelling bipoles is much greater than that at $1\arcsec$ resolution. The movie begins on June 29th, 2018 at 16:32:18 UT and ends the same day 18:38:36 UT. Its real-time duration is 14 seconds.

\bibliography{ReconnectionSolarWind_R1_ArXiv}{}
\bibliographystyle{apj}



\end{document}